\documentclass[preprint,showpacs,preprintnumbers,amsmath,amssymb]{revtex4}

\usepackage{graphicx}
\usepackage{dcolumn}
\usepackage{bm}
\newcommand {\apgt} {\ {\raise-.5ex\hbox{$\buildrel>\over\sim$}}\ }
\newcommand {\aplt} {\ {\raise-.5ex\hbox{$\buildrel<\over\sim$}}\ }

\begin{document}
 
\title{Infrared refractive index dispersion of PMMA spheres from 
synchrotron extinction spectra} 

\author{R. Bl\"umel$^1$, M. Ba\u{g}c{\i}o\u{g}lu$^2$, R. Lukacs$^2$, and A. Kohler$^2$}  
\affiliation{$^1$Department of Physics, Wesleyan University, 
Middletown, Connecticut 06459-0155}
\affiliation{$^2$Department of Mathematical 
Sciences and Technology, Faculty of Environmental 
Science and Technology, Norwegian University of Life Sciences, 1432 {\AA}s, Norway} 
  
\date{\today}

\begin{abstract} 
We performed 
high-resolution Fourier-transform infrared (FTIR) 
spectroscopy of a polymethyl methacrylate (PMMA) sphere 
of unknown size in 
the Mie scattering region. 
Apart from a slow, oscillatory structure (wiggles), which is 
due to an interference effect, the measured 
FTIR extinction 
spectrum exhibits a ripple structure, 
which is due to electromagnetic 
resonances. We fully characterize the underlying electromagnetic 
mode structure of the spectrum by 
assigning two mode numbers to each of the 
ripples in the measured spectrum. 
We show that analyzing the ripple structure in the spectrum 
in the wavenumber 
region from about $3000\,$cm$^{-1}$ to $8000\,$cm$^{-1}$ 
allows us to both determine the unknown radius of the sphere 
and the PMMA index of refraction, which shows 
a strong frequency dependence in this 
near-infrared spectral region. While in this paper 
we focus on examining a PMMA 
sphere as an example, our method of determining 
the refractive index and its dispersion 
from synchrotron infrared extinction spectra is 
generally 
applicable for the determination of the index of refraction 
of any transparent substance that can be shaped into micron-sized 
spheres. 
\end{abstract}



\maketitle
%

\section{Introduction} 
\label{Intro}
Although seemingly innocuous and analytically solvable within the 
theory of classical electromagnetism \cite{Jackson}, the 
scattering of light from dielectric spheres 
(``Mie scattering'' \cite{Mie,Debeye,vdH}) exhibits a 
multitude of interesting facets. For instance, 
it was not until the advent of modern digital computers 
(in this case the IBM 701, which became available 
in the early 1950s \cite{CH}) that the fully 
developed fine structure of 
the scattering cross section emerged \cite{Penn1}. 
Subsequently, this 
characteristic ``ripple structure'' \cite{vdH} 
gave rise to a lively debate 
on the physical origin of the ripples \cite{vdH,Chylek1,Chylek2} 
and led to some important applications, for instance the 
spectroscopy of ripples with the help of laser levitation 
\cite{Ashkin,Chylek3}. Even today, Mie scattering is important for 
applications ranging from light scattering of particles 
in the atmosphere \cite{AtPart} and colloidal solutions \cite{Mie,Coll} 
to its use as a reference system for infrared spectroscopy on 
single biological cells \cite{Bassan,BPhD}. It is the latter application 
that we have in mind when, in this paper, 
we study Mie scattering at PMMA 
spheres. 
While Fourier-transform infrared (FTIR) 
spectroscopy of PMMA spheres 
has been carried out in the wavenumber region up to 
$4000\,$cm$^{-1}$ \cite{Bassan,BPhD,Dijk}, 
one of the main points of this paper is to extend these 
measurements up to 
$8000\,$cm$^{-1}$, where the 
ripple patterns start to be fully developed. This serves as a test 
case for the resolving power of FTIR spectroscopy 
for applications to non-spherical systems, such as biological 
cells \cite{Bassan,McCann,Lasch}. 
We also present an algorithm that allows us to 
extract the radius and the index of refraction 
of micron-sized PMMA spheres
from 
the ripple structure of 
measured FTIR spectra. 
This extends our knowledge of the PMMA index of 
refraction into the near-infrared region from $3000\,$cm$^{-1}$ to 
$8000\,$cm$^{-1}$, where 
literature data, to our knowledge, are absent. 
It also allows us 
to test and challenge extrapolation formulae for the 
index of refraction in this range of wavenumbers 
\cite{BORN,CAUCHY,Sell,Kasa1}. 
In general, we find that FTIR 
spectroscopy of PMMA spheres is an excellent technique to 
establish the limits of FTIR spectroscopy, in particular as far as resolution 
is concerned. It may be used for calibration of the 
FTIR equipment before spectra of biological samples are taken. 
From the theoretical point of view our experimental extinction spectra 
of synchrotron radiation on PMMA microspheres test the Mie theory of 
scattering on dielectric spheres. We accomplish this by 
classifying each ripple 
in the extinction curve of PMMA micro spheres with electromagnetic 
mode numbers 
that uniquely characterize the specific nature and physical origin 
of each individual ripple in the spectrum.
Our methods may be applied to high-resolution FTIR spectroscopy 
of micro-spheres of any transparent biological or 
inanimate material. In particular, we
propose to use high-resolution FTIR spectroscopy in conjunction with 
our methods as a new spectroscopic tool for the determination 
of the index of refraction of transparent materials in the near- to far-infrared 
spectral regime, where FTIR spectroscopy is conventionally performed. 
 
 
\section{Theory} 
\label{THEO} 
To set the stage for the analysis of our synchrotron FTIR spectra, 
we present in this section some background material on 
the scattering of infrared radiation from non-absorbing 
dielectric spheres 
of radius $R$, geometric cross section $g=\pi R^2$, real 
refractive index $n$, and magnetic permeability $\mu=1$. 
These assumptions are appropriate for the analysis of the 
experiments described in this paper in which we focus on 
scattering of infrared radiation at PMMA spheres ($\mu\approx 1$) in 
the spectral region from $3000\,$cm$^{-1}$ to $8000\,$cm$^{-1}$  
in which there is very little absorption 
($n$ approximately real). 
In our FTIR synchrotron experiments, an infrared beam 
with intensity $I_0$ is incident on a dielectric sphere, where it 
may be scattered, absorbed, or transmitted into 
a detector with surface area $G>g$. 
The scattered intensity is denoted by $I_{\rm sca}$, 
the absorbed intensity by $I_{\rm abs}$, and the un-scattered 
intensity, directed strictly in forward direction, 
is denoted 
by $I$. The cross sections for scattering and 
absorption are denoted by 
$\sigma_{\rm sca}$ and $\sigma_{\rm abs}$, respectively. 
We also define the extinction cross section 
$\sigma_{\rm ext} = \sigma_{\rm sca}+\sigma_{\rm abs}$ 
\cite{vdH}. 
With the help of $G$ and the cross sections we may 
compute the associated radiative powers 
$P_0=I_0 G$, 
$P_{\rm sca}=I_{0}\sigma_{\rm sca}$, 
$P_{\rm abs}=I_{0}\sigma_{\rm abs}$, and 
$P=IG$. 
Conservation of power 
requires: 
\begin{equation}
P_0 = P + P_{\rm sca} + P_{\rm abs} . 
\label{THEO1}
\end{equation}
Therefore, from (\ref{THEO1}), we obtain 
\begin{equation}
I_0 G = I G + I_0\sigma_{\rm sca} + I_0\sigma_{\rm abs} 
          = IG + I_0\sigma_{\rm ext}. 
\label{THEO2}
\end{equation}
In our experiments we measure the {\it apparent absorbance}, 
defined as 
\begin{equation}
A = - \log_{10} \left(\frac{I}{I_0}\right). 
\label{LOOK1}
\end{equation}
This quantity is called the apparent absorbance since it does not 
only include the intensity lost due to (chemical) absorption 
(i.e., the ``true'' absorption), but also the intensity lost via 
scattering. 
Because of the finite area $G$ of our detector, and because 
of its finite distance to the scatterer, apart from collecting 
the un-scattered intensity $I$, our detector also collects some 
scattered light. Therefore, the recorded intensity $I$ is 
somewhat different from the un-scattered intensity in 
forward direction. 
However, since the scattered intensity reaching our detector 
is small, and since, in addition, we are mainly interested 
in the {\it structure} of the resulting extinction spectra, 
not absolute values of the 
extinction, we found this effect to be negligible. 
  
Following \cite{vdH}, we define the extinction efficiency 
\begin{equation}
Q_{\rm ext} = \frac{\sigma_{\rm ext}}{g}. 
\label{THEO3}
\end{equation}
To convert the results of our absorbance measurements to 
$Q_{\rm ext}$, we proceed as follows.
From (\ref{LOOK1}) we obtain 
\begin{equation} 
\left( \frac{I}{I_0}\right) = 10^{-A}. 
\label{LOOK1a}
\end{equation}
Dividing (\ref{THEO2}) by $I_0 G$ and 
using (\ref{THEO3}) and (\ref{LOOK1a}), we obtain 
\begin{equation}
1 = \frac{I}{I_0} + \frac{\sigma_{\rm ext}}{G} 
= 10^{-A} + \frac{g}{G} Q_{\rm ext},
\label{LOOK1b}
\end{equation}
from which we obtain 
\begin{equation} 
Q_{\rm ext} = \left(\frac{G}{g}\right) \left( 1 - 10^{-A} \right).
\label{LOOK2}
\end{equation}
For dielectric spheres 
the extinction efficiency 
$Q_{\rm ext}$ can be computed analytically. 
This was first accomplished by Mie in 1908 \cite{Mie} 
and further developed by Debeye in 1909 \cite{Debeye}. 
In the modern literature the result is usually quoted either in 
the notation of van de Hulst \cite{vdH} or Newton \cite{Newton}. 
In the notation of van de Hulst we have 
%
\begin{equation}
Q_{\rm ext} = \frac{2}{x^2} \sum_{n=1}^{\infty} 
(2n+1) \Re(a_n+b_n), 
 \label{THEO4}
\end{equation}
where $a_n$ and $b_n$ denote the complex-valued 
Mie coefficients \cite{vdH}, $\Re$ denotes the real part, 
\begin{equation}
x=kR = 2\pi R/\lambda = 2\pi R \tilde \nu
 \label{THEO5}
\end{equation}
is the size parameter, $\lambda$ is the vacuum 
wavelength, $\tilde \nu=1/\lambda$ is the wavenumber, 
and 
$k=2\pi/\lambda=2\pi\tilde\nu$ is the angular wavenumber. 
In the 
notation of Newton \cite{Newton} we have 
\begin{equation}
Q_{\rm ext} = \frac{1}{x^2} \sum_{J=1}^{\infty} 
(2J+1)\left(2-\Re S_e^J - \Re S_m^J\right),
 \label{THEO6}
\end{equation}
where $S_e^J$ and $S_m^J$ are the scattering matrix 
($S$-matrix) elements for electric and magnetic 
multipole radiation of order $2^J$, respectively. Explicitly, 
$S_e^J$ and $S_m^J$ are given by 
\begin{equation} 
S_e^J(x;n) = e^{i\pi J}\ \frac{n{w_J^{(-)}}'(x) u_J(nx) - w_J^{(-)}(x)  u'_J(nx)} 
{n{w_J^{(+)}}'(x) u_J(nx) - w_J^{(+)}(x) u'_J(nx)} 
 \label{THEO7}
\end{equation}
and 
\begin{equation} 
S_m^J(x;n) = e^{i\pi J}\ \frac{{w_J^{(-)}}'(x)u_J(nx) - nw_J^{(-)}(x) u'_J(nx)} 
{{w_J^{(+)}}'(x) u_J(nx) - nw_J^{(+)}(x) u'_J(nx)} , 
\label{THEO8}
\end{equation}
where $J\geq 1$ and 
$u_J(z)$ and $w_J^{(\pm)}(z)$ are 
defined with the help of the spherical Bessel functions $j_J(z)$ 
and $n_J(z)$ \cite{Newton,AS} according to 
\begin{align}
u_J(z) &= z j_J(z),\ \ \ v_J(z) = z n_J(z),\ \ \ w_J^{(+)}(z) = 
e^{i\pi(J+1)} [v_J(z) - i u_J(z)], 
\nonumber \\ 
w_J^{(-)}(z) &= w_J^{(+)}(-z) = e^{i\pi J} w_J^{(+)}(z) ^* , 
\label{THEO9}
\end{align}
where the star indicates complex conjugation and 
the prime in (\ref{THEO7}) and (\ref{THEO8}) indicates 
differentiation with respect to the argument. 
 
The representations (\ref{THEO4}) and (\ref{THEO6}) of 
$Q_{\rm ext}$ are identical. The connection is 
established via 
\begin{equation} 
a_n = \frac{1}{2}\left(1-\Re S_e^n\right),\ \ \ 
b_n = \frac{1}{2}\left(1-\Re S_m^n\right).  
\label{THEO9a} 
\end{equation}
While 
(\ref{THEO4}) and 
(\ref{THEO6}) are equivalent, 
Newton's notation in (\ref{THEO7}) and (\ref{THEO8}) 
emphasizes the physical meaning of the input 
quantities in $Q_{\rm ext}$ as the two types of radiation, 
namely electric and magnetic multi-pole radiation \cite{Jackson}. 
This is convenient, in particular  
for a classification of the features observed in $Q_{\rm ext}$ 
as a function of $x$. Therefore, for the rest of this paper, 
we will adhere to Newton's notation \cite{Newton}. The expansions 
(\ref{THEO4}) and (\ref{THEO6}) are known as partial-wave 
expansions, since each term in the expansion corresponds to 
a specific scattering mode. This is most prominently exhibited 
by (\ref{THEO7}) and (\ref{THEO8}),
where $S_e^J$ is the amplitude for scattering into the 
electric partial wave represented by the transverse 
vector spherical harmonic $Y^{(e)}_{JM}$ and 
$S_m^J$ is the amplitude for scattering into the 
magnetic partial wave represented by the transverse 
vector spherical harmonic $Y^{(m)}_{JM}$, where 
the indices $e$ and $m$ stand for electric and magnetic, 
respectively, and $M$, an integer, ranges from $-J$ to $J$ 
in steps of 1.   
  
%
\begin{figure}
\includegraphics{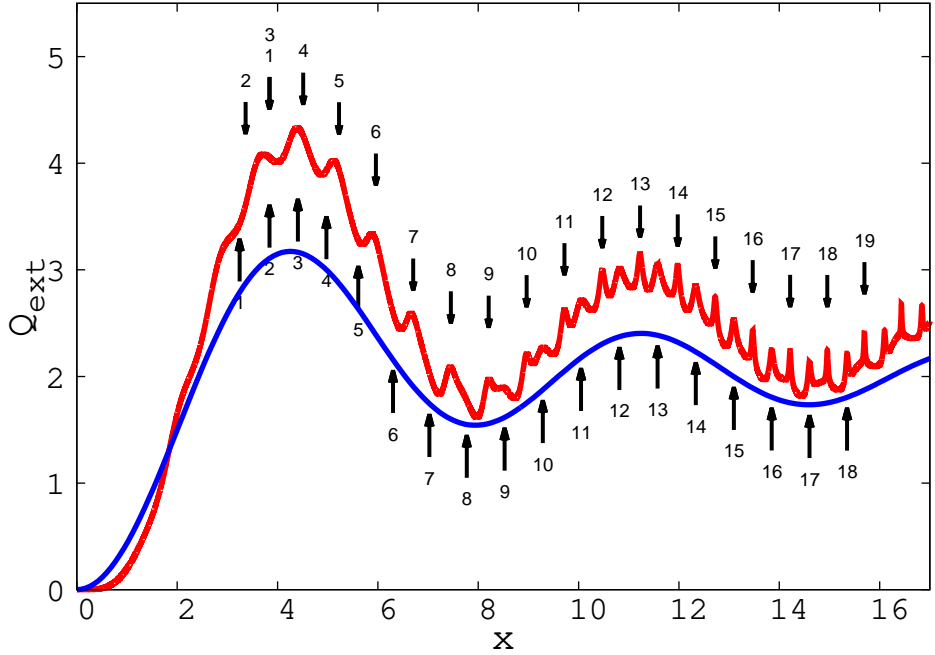}
\caption{\label{Fig1} (Color online) Red, solid line: 
Extinction efficiency $Q_{\rm ext}$ as a function 
of size parameter $x$ for $n=1.48$. A ripple structure is observed,  
superimposed 
on a wavy background. As $x$ increases, the ripples become 
sharper. Each ripple is associated with a partial-wave resonance 
in either the electric ($S_e^J$) or magnetic ($S_m^J$) 
S-matrix elements. Downward arrows point to the 
magnetic resonances while upward arrows point to the 
electric resonances. The $J$ classification of magnetic 
and electric ripples is 
shown above and below the tails of the respective arrows. 
The magnetic peaks $J=1$ and $J=3$ are nearly degenerate and 
also nearly degenerate with the electric 
$J=2$ peak. Blue, solid line: Analytical van de Hulst 
approximation of $Q_{ext}$ according to (\ref{THEO11}). 
 } 
\end{figure}
%
 
At this point we have assembled all the tools necessary to plot 
the extinction efficiency $Q_{\rm ext}$ 
as a function of $x$. The red curve in 
Fig.~\ref{Fig1} shows $Q_{\rm ext}$, calculated according to 
(\ref{THEO6}), 
in the interval $0<x<17$ for $n=1.48$,  
the approximate index of refraction of PMMA in the transition 
region between the optical and the near infrared \cite{Kasa1}. 
We see that 
$Q_{\rm ext}$ 
exhibits the following three distinctive features.  
\begin{enumerate} 
\item {\bf Long-range oscillations:} This feature 
is explained by an interference effect. 
The phase lag between the central ray passing through 
a sphere of refractive index $n$ compared to the same ray in the 
absence of the refracting sphere is 
\begin{equation}
\rho = 2x(n-1) . 
\label{THEO10}
\end{equation}
This phase lag may be used to derive an approximate analytical 
expression for the long-range oscillations in $Q_{\rm ext}(x)$, first 
obtained and published by van de Hulst \cite{vdH}, 
\begin{equation}
Q_{\rm ext}(x) = 2 - \frac{4}{\rho} \sin(\rho) + 
\frac{4}{\rho^2} [1-\cos(\rho)] . 
\label{THEO11}
\end{equation}
It is shown as the blue solid line in Fig.~\ref{Fig1}. 
The fit is not perfect, but captures the frequency of 
the long-range oscillations very well. In particular, the 
fit is convincing enough to indicate that the basic 
physical origin for 
the long-range oscillations as an interference effect 
is properly captured. 
\item {\bf Ripples:} This feature, a fine structure of peaks 
superimposed on the 
long-range oscillations in $Q_{\rm ext}$, is due to 
partial-wave resonances in the S-matrix elements 
($S_e^J$) and ($S_m^J$) defined in 
(\ref{THEO7}) and (\ref{THEO8}), respectively
\cite{Chylek1,Chylek2}. The resonances are 
called ripples in \cite{vdH}. 
Since the ripples form the basis for our technique of extracting the 
radius and index of refraction of dielectric spheres from 
synchrotron FTIR spectra, we will discuss the ripples 
in more detail below. 
\item {\bf Extinction Paradox:} As shown in 
Fig.~\ref{Fig1}, $Q_{\rm ext}$ oscillates around 
a value in the vicinity of $2$, about twice 
the classically expected $Q_{\rm ext}=1$.  
This is known as the 
extinction paradox (see, e.g., \cite{ep}), 
resulting from a combination of scattering and 
diffraction, 
widely discussed and explained in the literature 
(see, e.g., \cite{vdH,Newton,ep}).  
\end{enumerate} 
%
 
%
\begin{figure}
\includegraphics{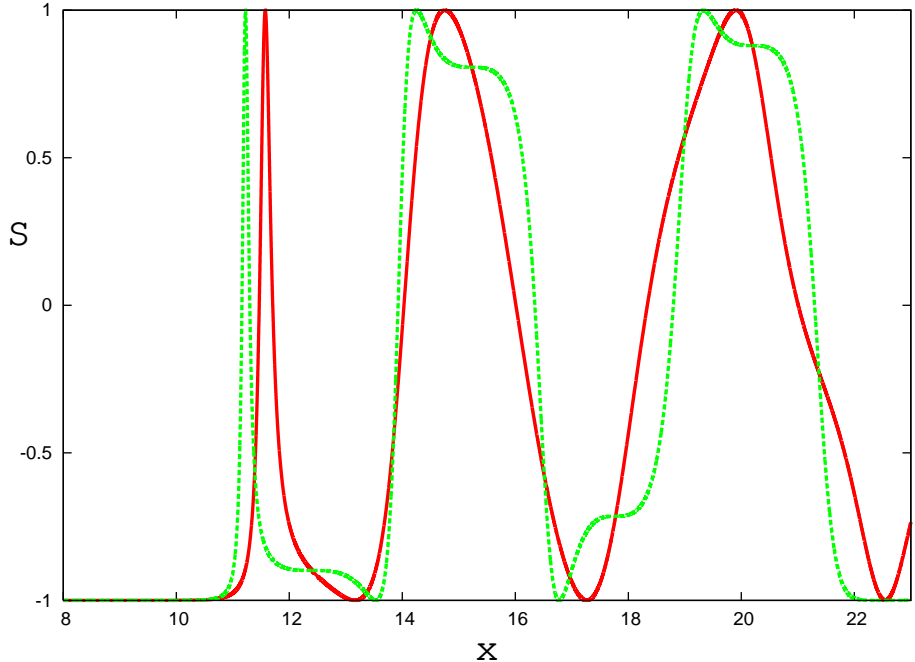}
\caption{\label{Fig2} (Color online) 
$S$-matrix elements 
$S_e^{J=13}$ (red, solid line) and $S_m^{J=13}$ (green, dashed line) as a function 
of size parameter $x$ for $n=1.48$. Both S-matrix elements exhibit 
an infinite series of resonances; the first three of them are shown in the figure. 
The first resonance, in both the electric and magnetic 
S-matrix elements, is sharp, followed by broader 
resonances for larger $x$. The first (sharp) magnetic peak occurs at 
$x=11.2245$; the first (sharp) electric peak occurs at $x=11.5750$. 
 } 
\end{figure}
%
 
The resonance structures in $Q_{\rm ext}(x)$, 
i.e., the ripples (see Fig.~\ref{Fig1}), are due to 
the complex zeros of the denominators of the S-matrix elements 
$S_e^J$ and $S_m^J$ in (\ref{THEO7}) and (\ref{THEO8}), 
respectively, 
which correspond to 
poles of the S-matrix elements in the complex $x$ plane. 
In general, 
if the S-matix 
poles are far from the real $x$ axis, i.e., they have 
a large imaginary part, 
the corresponding 
resonances are wide; if the poles are close to the real $x$ 
axis (small imaginary part), 
the corresponding resonances are sharp \cite{Newton}. 
Apparently, as shown in Fig.~\ref{Fig1}, in the case 
of Mie scattering, the resonances, i.e., the ripples, 
are getting sharper with increasing $x$. Electromagnetic 
waves with fixed $J$ are also known as partial waves 
\cite{Newton}. Therefore, the ripples in Fig.~\ref{Fig1} 
are partial-wave resonances \cite{Chylek1,Chylek2,Newton}. 
 
For each $J$ there is an infinite sequence of partial-wave 
resonances that occur at positions 
denoted by $x_{J,p}$, 
$p=1,2,\ldots$, ordered such that $x_{J,p+1}>x_{J,p}$. 
This is illustrated in Fig.~\ref{Fig2}, which shows the 
S-matrix elements $S_e^{J=13}$ and $S_m^{J=13}$ 
as a function of $x$ for $n=1.48$. 
 
In order to classify the ripple (resonance) structure 
in $Q_{\rm ext}$, we 
introduce the notation 
$ (J,M;e)$ und $(J,M;m)$ to denote the 
electromagnetic modes corresponding to 
electric and magnetic $2^J$-pole radiation, 
respectively. For instance, the mode 
$(1,M;e)$ corresponds to electric dipole radiation 
\cite{Jackson} where $M$ may assume the values 
$-1,0,+1$. For a spherical scatterer 
the resonance positions depend only on $J$; 
they are degenerate in $M$. 
 
%
\begin{figure}
\includegraphics{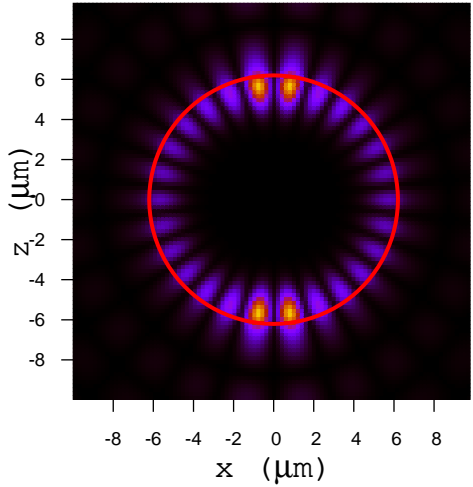}
\caption{\label{Fig3} 
(Color online) 
Heat map of the absolute square 
of the 
electric field of the $J=13$, $M=0$ magnetic 
mode $(J=13,M=0;m)$ at $x=11.224$ 
in the $x$-$z$ equatorial plane of a dielectric sphere 
with radius $R=6.2\,\mu$m and index of refraction $n=1.48$. 
The brighter the color, the larger the magnitude of 
the electric field. 
Also shown (red solid line) is the intersection of the 
surface of the sphere and the equatorial $x$-$z$ plane.  
 } 
\end{figure}
%
 
According to \cite{Chylek1} the spacings between resonances 
of a given 
mode do not depend on the nature of this 
mode (electric or magnetic) and are given approximately by 
\cite{Chylek1}
\begin{equation}
\Delta x(n) = \frac{\arctan \left[ \left(n^2 - 1\right)^{1/2} \right] } 
{\left( n^2 - 1\right)^{1/2}} . 
\label{THEO12}
\end{equation}
This shows that the spacing between resonances is approximately 
independent of $x$ and depends only on the refractive index $n$. 
Therefore, we may use the spacings between experimentally 
observed ripples to determine the index of refraction and its 
dependence on the wavelength (dispersion). 
 
Formula (\ref{THEO12}) is reasonably accurate. 
For $n=1.48$, e.g., and in the wavenumber range 
of interest in this paper ($3000\,$cm$^{-1}$,\ldots,
$8000\,$cm$^{-1}$), (\ref{THEO12}) 
is accurate to 
within about 2\%. In addition, for $n=1.48$ and $J$ in the 
vicinity of 13, it is even accidentally exact. 
While a 2\% accuracy is sufficient for rough estimates, 
a precision determination of the index of refraction from 
synchrotron FTIR spectra requires better accuracy. 
In this case we need to resort to the exact computation 
of $\Delta x$ via (\ref{THEO6}), (\ref{THEO7}), and 
(\ref{THEO8}). 
 
From (\ref{THEO12}) we obtain 
\begin{equation}
\frac{d \Delta x(n)}{dn} = -\left[ \frac{n}{(n^2-1)^{3/2}} \right] 
\arctan\left[ (n^2-1)^{1/2}\right] + 
\frac{1}{n(n^2-1)}. 
\label{THEO13}
\end{equation}
At $n=1.48$ we obtain $d\Delta x/dn = -0.38$. This shows that 
the index of refraction $n$ reacts reasonably sensitively to 
changes in $\Delta x$, a prerequisite for 
using ripple spacings for determining the index of refraction. 
 
 
In order to illustrate the mode structures of the resonances 
in Fig.~\ref{Fig1}, we show in 
Fig.~\ref{Fig3}, as an example, the electric-field 
distribution of the $J=13$, $M=0$ magnetic 
mode $(J=13,M=0;m)$ that occurs at $x=11.224$ in Fig.~\ref{Fig1}. 
Shown is 
the absolute square, 
$|\vec E|^2=\vec E\cdot \vec E^*$ (the star 
denotes complex conjugation), 
of the 
electric field of the mode 
in the $x$-$z$ equatorial plane of a dielectric sphere 
with radius $R=6.2\,\mu$m and index of refraction $n=1.48$. 
We see that in analogy to 
a whispering-gallery mode in acoustics 
\cite{Rayleigh} or dielectric 
microresonators \cite{Chiasera}, 
the electric  
field strength is guided along the surface of the sphere 
and is maximal just inside of the sphere, very close to 
the sphere's surface. Since total internal reflection 
accounts for the confinement of this wave, the quality 
factor $Q$ for these modes is extraordinarily large. 
Counting the maxima of the field in Fig.~\ref{Fig3}, we obtain 
26, which is $2J$. This is consistent, since 
in Fig.~\ref{Fig3} we plot the absolute square of 
$\vec E$, which produces two 
maxima per wavelength. 
 
As shown in Fig.~\ref{Fig1}, 
the resonances corresponding to the magnetic modes 
are about a factor 2 sharper than the resonances corresponding 
to electric modes. 
Therefore, in the following, we will 
concentrate on analyzing the magnetic modes exclusively. 
The sharpness of the magnetic resonances also gives us 
better resolution, which helps greatly in assigning 
the correct mode numbers to the ripples in 
FTIR spectra.

 
\section{Experiment} 
\label{FSE} 
To test the theory and in order to find out whether 
synchrotron-based FTIR spectroscopy is powerful enough and 
has enough resolution to exhibit ripples in the 
extinction spectrum, we measured the extinction spectrum 
of a polymethyl methacrylate 
(PMMA) microsphere taken from a sample of 
PMMA microspheres, purchased from 
MicrospheresÐNanospheres (Corpuscular Inc, NY). 
The sample was labeled by the manufacturer as consisting 
of plain PMMA spheres with an average diameter of $5.5\,\mu$m. 
The spread of diameters in the sample was not given. 
From this sample we selected a sphere at random and 
used it in our measurements without any 
modifications. 
 
%
\begin{figure}
\includegraphics[scale=1.3]{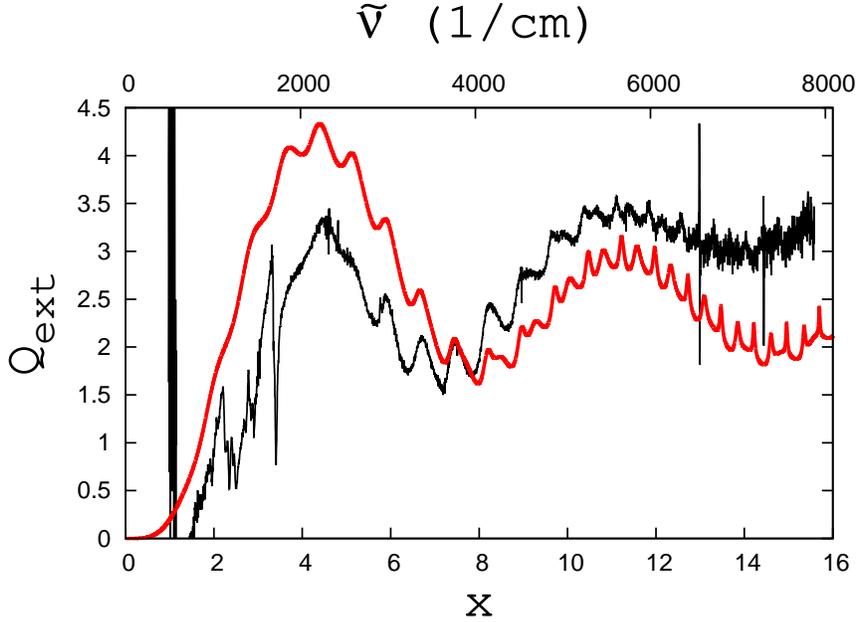}
\caption{\label{Fig4} (Color online) 
$Q_{\rm ext}$ for the experimentally measured sphere versus 
the size parameter $x$ (bottom scale). The top scale shows 
the wavenumber $\tilde\nu$ in 1/cm. The conversion between 
$x$ and $\tilde\nu$ is accomplished by assuming a sphere 
radius of $3.15\,\mu$m. 
Black, solid line: Experimental result of our 
synchrotron infrared absorption measurements. 
Red, solid line: 
Theoretical extinction efficiency imported from Fig.~\ref{Fig1}. 
The experimental extinction curve shows both wiggles 
(long-range oscillations) and ripples 
(sharp partial-wave resonances). The features 
in the experimental extinction spectrum are seen to line 
up with corresponding features in the theoretical spectrum. 
 } 
\end{figure}
%
 
In order to obtain a high-quality spectrum for this sphere, 
we used synchrotron 
radiation provided by the MAX III synchrotron facility 
in Lund, Sweden. The $Q_{\rm ext}$ spectrum of the sphere 
was recorded with a resolution of $2\,$cm$^{-1}$ in the 
wavenumber range from $650\,$cm$^{-1}$ to 
$8000\,$cm$^{-1}$ 
as an average consisting of 256 
individual scans with a 
Bruker Hyperion 3000 IR microscope (Bruker Optik, Germany), 
coupled with an FTIR spectrometer (Bruker IFS66V),  
equipped with a liquid-nitrogen-cooled 
single-element mercury cadmium telluride (MCT) 
$100\,\mu{\rm m}\times 100\,\mu{\rm m}$ detector.
We used a 
$15\times$ objective with an aperture size of 
$10\,\mu$m$\times 10\,\mu$m. 
As a substrate, $3\,$mm-thick ZnSe 
spectrophotometric optical slides were used, 
and a clean ZnSe plate was used as a reference. 
 
The raw experimental $A(\tilde\nu)$ spectrum was 
converted to 
$Q_{\rm ext}(\tilde\nu)$ via (\ref{LOOK2}). 
Since we are interested in the {\it structure} 
of the experimental $Q_{\rm ext}(\tilde\nu)$
spectrum, not the absolute magnitude, no 
attempt was made to obtain an accurate 
value for the factor $f=G/g$ in 
(\ref{LOOK2}). Instead we use this factor as a 
scale factor to shift the experimental 
$Q_{\rm ext}(\tilde\nu)$ spectrum 
into the vicinity of 
the theoretical $Q_{\rm ext}(\tilde\nu)$ spectrum 
for convenient comparison. Since we are working with 
a sphere, which nominally has a diameter of $5.5\,\mu$m, 
and since our aperture is $10\,\mu$m$\times 10\,\mu$m, 
a good estimate for $f$ is 
$f=10\,\mu$m$\times 10\,\mu$m/$[\pi\times (2.75\,\mu$m$)]^2=4.2$. 
We chose $f=5$ for a good visual presentation of 
the data. 
 
Although we already see ripples in the raw spectrum, 
the raw $Q_{\rm ext}(\tilde\nu)$ data 
are noisy and many of the ripples are 
hidden in the noise. 
To bring them out, we 
performed a running average 
on the raw data according to 
\begin{equation}
\bar Q_{\rm ext}(\tilde\nu) = \frac{1}{21} 
\sum_{j=-10}^{10} Q_{\rm ext}(\tilde\nu+j s )
\label{EXPAV}
\end{equation}
where $s=0.964\,$cm$^{-1}$ is the wavenumber step-size in 
our experiments. 
The resulting 
averaged experimental spectrum with $f=5$ 
is shown as the black solid line in 
Fig.~\ref{Fig4}. Ripples in the 
$\bar Q_{\rm ext}(\tilde\nu)$ spectrum 
are now clearly visible. 
Also shown is the theoretical $Q_{\rm ext}(\tilde\nu)$ 
spectrum (red, solid line) that we imported into 
Fig.~\ref{Fig4} directly from Fig.~\ref{Fig1} without 
any modifications. We see that both wiggles and 
ripples in the experimental spectrum line up 
very well with the corresponding structures in 
the theoretical spectrum. No agreement 
between the experimental $\bar Q_{\rm ext}(\tilde\nu)$ 
spectrum and the corresponding theoretical 
spectrum 
is expected for $\tilde\nu<3000\,$cm$^{-1}$, 
since PMMA absorbs strongly in this wavenumber 
region. Therefore, we focus in this paper 
on the wavenumber region $\tilde\nu > 3000\,$cm$^{-1}$. 
 
Table~\ref{TAB1} lists the positions of 14 sharp,  
consecutive 
ripples, extracted 
from the averaged experimental spectrum 
in Fig.~\ref{Fig4}. These ripples correspond 
to $p=1$ magnetic modes, whose angular momenta 
$J$ were assigned unambiguously 
(see Table~\ref{TAB1}). The first resonance 
listed in Table~\ref{TAB1} ($J=6$) 
is the first resonance to the right of the 
absorption region. 

\begin{table}[ht] 
\caption{\label{TAB1} Locations $\tilde \nu$ 
(in cm$^{-1}$)
of experimental $p=1$ magnetic Mie resonances 
extracted from 
the experimental FTIR spectrum 
(black, solid line in Fig.~\ref{Fig4})  
for the measured PMMA sphere.
The error in the listed wavenumbers, due to uncertainty 
in determining the peak locations, 
is approximately $\pm 5\,$cm$^{-1}$. 
} 
\centering      
\begin{tabular}{|| c  | c || c | c || c | c ||}  
\hline        
$J$ & $\tilde\nu$  & $J$ & $\ \tilde\nu\ $ & $\ J\ $ & $\tilde\nu$  \\ [0.5ex] 
\hline\hline                     
6   &  3028  &   11  &  4950  &  16 &   6822     \\
7   &  3450  &   12  &  5338  &  17 &   7204     \\  [1ex]       
8   &  3827  &   13  &  5703  &  18 &   7581    \\
9   &  4233  &   14  &  6078  &  19 &   7962     \\
10  & 4589  &   15  &  6448  &     &       \\
\hline     
\end{tabular} 
\label{Tab1}  
\end{table} 
%

\section{PMMA Index of Refraction: Known Results} 
\label{PMMAIDX}
 
%
\begin{figure}
\includegraphics{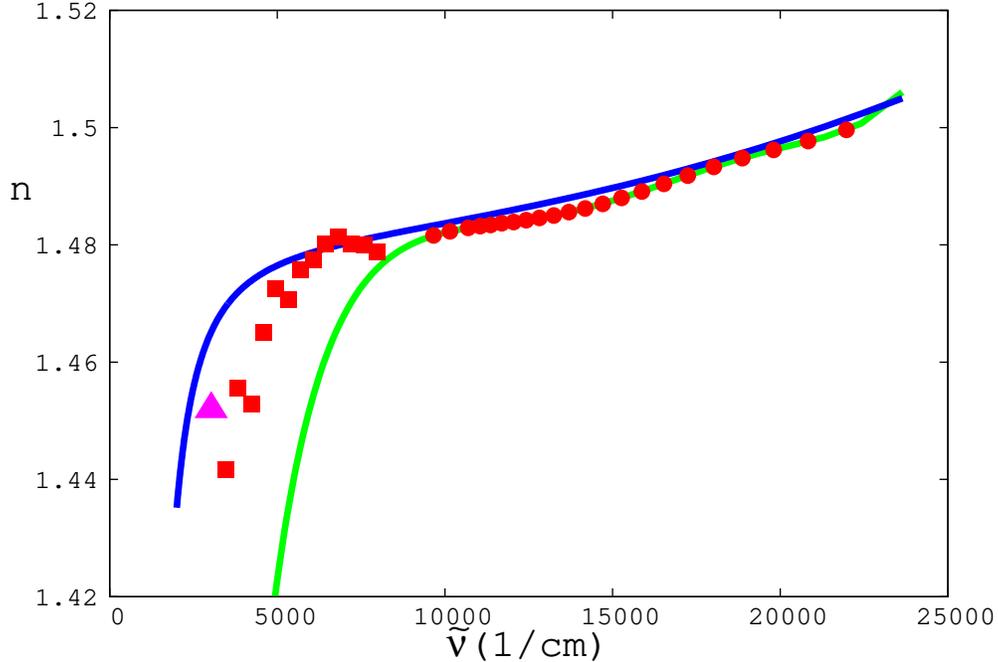}
\caption{\label{Fig5} (Color online) 
PMMA index of refraction vs. wavenumber. 
Red plot symbols: Experimental data according to \cite{Kasa1}. 
Green line: Fit of the experimental data points with an extended 
Cauchy formula \cite{Kasa1}. Blue line: 
Three-term Sellmeier fit of the PMMA index 
of refraction according to \cite{Ishi}. 
Red, filled squares: Index of refraction as a function of wavenumber 
extracted from our experimental FTIR synchrotron data. 
Magenta, filled triangle: Data point for $J=6$. As an outlier it 
is characterized by a different plot symbol. 
 } 
\end{figure}
%
Experimental data for the PMMA refractive index 
in the mid to far IR regions 
are not available. 
Therefore, based on data available in 
the optical and near IR regime,
we need to extrapolate. 
Two extrapolation methods are 
available: Cauchy and Sellmeier. 
%
%

Cauchy's formula \cite{BORN}, first published in the 
1830s \cite{CAUCHY}, is given explicitly by 
\begin{equation}
    n(\lambda) = B + \frac {C}{\lambda^2} + \frac{D}{\lambda^4} + \cdots, 
\label{IDX1}
\end{equation}
where $B$, $C$, $D$, $\ldots$ are fit parameters and $\lambda$ 
is the vacuum wavelength of the incident light. An extended 
Cauchy formula, keeping terms up to 
8th order in (\ref{IDX1}), was fitted by the authors of \cite{Kasa1}. 
The result is shown as the green line in Fig.~\ref{Fig5}. 
Apparently, in 
the optical and near-infrared regimes the fit is excellent. 
However, Cauchy's formula is known 
to be unreliable outside the fit region and should not be used 
to extrapolate too far into the mid- and far-infrared regions. 

Sellmeier's formula \cite{BORN}, first published in 1871 
\cite{Sell}, is explicitly given by  
\begin{equation}
n(\lambda) = \left[ 1+ \sum_j \frac{A_j \lambda^2}{\lambda^2 - B_j^2} \right] ^{1/2}. 
\label{IDX2}
\end{equation}
A three-term Sellmeier formula for PMMA is fitted in \cite{Ishi}. 
These authors quote: 
$A_1$ = 0.4963, $B_1=71.80$nm, 
$A_2=0.6965$, $B_2=117.4$nm, and 
$A_3=0.3223$, $B_3=9237$nm. 
The result is shown as the blue line in Fig.~\ref{Fig5}. 
It is known that, compared with Cauchy's formula, 
Sellmeier's formula is considerably more reliable 
for extrapolation into the mid- and far-infrared regions. 
This is corroborated by the following argument. 
According to Table~\ref{TAB1}, the spacing $\Delta\tilde\nu$ 
between magnetic resonances is nearly constant from 
$J=13$ to $J=19$, corresponding to wavenumbers 
ranging from $\tilde\nu=5703\,$cm$^{-1}$ to 
$\tilde\nu=7962\,$cm$^{-1}$. This indicates that, 
according to (\ref{THEO12}), the index of refraction 
of our PMMA sphere in this wavenumber range 
is nearly constant. This is inconsistent with 
the strong wavenumber dependence of 
the Cauchy extrapolation formula in this 
wavenumber range (see Fig.~\ref{Fig5}) and 
suggests that, as expected, 
the Sellmeier extrapolation formula in this spectral 
regime is more reliable than the Cauchy formula.

\section{Sphere Radius} 
\label{SR}
In this section we describe several methods that 
are capable of determining 
the radius of an isolated dielectric sphere and its index of refraction 
from measured (FTIR) extinction data. While the methods described 
are applicable in general to the determination of the radius of 
any transparent, dielectric sphere,
we will apply these methods, as a specific 
example, 
to the determination 
of the radius of our measured PMMA sphere. 

Our first method is purely analytical and yields a rough 
analytical estimate of the radius $R$ of our measured PMMA 
sphere. 
For constant $R$ and corresponding 
$x$ and $\tilde\nu$ intervals, $\Delta x$ and $\Delta\tilde\nu$, 
respectively, (\ref{THEO5}) 
implies
\begin{equation}
\Delta x = 2\pi R \Delta\tilde\nu , 
\label{STIDX1a}
\end{equation}
which may be solved for $R$ to yield 
\begin{equation}
R=\frac{\Delta x}{2\pi\Delta\tilde\nu}.
\label{SRIDX0}
\end{equation}
As pointed out in Sec.~\ref{PMMAIDX}, 
in the wavenumber interval from 
$J=13$ ($5703\,$cm$^{-1}$) to 
$J=19$ ($7962\,$cm$^{-1}$)
(see Table~\ref{TAB1} for the assignment of wavenumbers to $J$ values), 
the spacing $\Delta\tilde\nu$ 
between magnetic resonances is nearly constant. 
Since there are six 
spacings in this wavenumber interval, the average 
spacing is 
\begin{equation}
\Delta\tilde\nu = (7962\,{\rm cm}^{-1} - 5703\,{\rm cm}^{-1}) / 6 = 
377\,{\rm cm}^{-1}. 
\label{SRIDX2}
\end{equation}
We also argued in Sec.~\ref{PMMAIDX} that 
Sellmeier's formula is realistic in this wavenumber interval. 
It predicts 
an index of refraction of $n\approx 1.48$, for which 
(\ref{THEO12}) yields 
$\Delta x=0.760$. Using this value of $\Delta x$ 
together with (\ref{SRIDX2})
in (\ref{SRIDX0}), we obtain 
\begin{equation}
R =  3.21\,\mu{\rm m}. 
\label{SRIDX3}
\end{equation}

Instead of using the analytical formula (\ref{THEO12}) for 
the average spacing $\Delta x$, our second method 
uses the exact spacing derived from the exact peaks 
in $Q_{\rm ext}(x)$. For $n=1.48$ the 
magnetic $J=13$ peak in the exact 
$Q_{\rm ext}(x)$ occurs at $x=11.2225$, while the peak 
at $J=19$ occurs at $x=15.6959$. This implies an 
average spacing of $(15.6959-11.2225)/6=0.746$, which, 
used together with (\ref{SRIDX2}) in (\ref{SRIDX0}), yields 
\begin{equation}
R = 3.15\,\mu{\rm m}. 
\label{SRIDX4}
\end{equation}

We may also base our radius estimate on the $J=19$ ripple itself. 
Starting directly with (\ref{THEO5}), we have: 
\begin{equation}
R=\frac{x_{J=19}}{2\pi\tilde\nu_{J=19}} = \frac{15.70}{2\pi\, 7962\,{\rm cm}^{-1}} 
= 3.14\,\mu{\rm m}
\label{SRIDX4a}
\end{equation}
All three results, (\ref{SRIDX3}), (\ref{SRIDX4}), 
and (\ref{SRIDX4a}) are consistent within about $\pm 35\,$nm. 

In order to obtain a rough analytical estimate of the expected 
uncertainty $\Delta R$ in $R$, we 
allow an error of 
$\delta\Delta\tilde\nu=5\,$cm$^{-1}$ in 
$\Delta\tilde\nu$ and an error of 
$\delta n = 0.002$ in $n$ and 
compute $\Delta R$ according to 
\begin{align}
\Delta R &= \left[ \left( \frac{\partial R}{\partial n}\delta n\right)^2 
+\left( \frac{\partial R}{\partial \Delta\tilde\nu} \delta\Delta\tilde\nu \right) ^2 
\right] ^{1/2} \\ 
&=R\left[ \left(\frac{1}{\Delta x} \frac{d \Delta x}{d n}\delta n\right)^2 
+\left( \frac{\delta\Delta\tilde\nu}{\Delta\tilde\nu}  \right) ^2 
\right] ^{1/2},  
\label{SRIDX5}
\end{align}
where we used (\ref{SRIDX0}). Using the 
analytical expression (\ref{THEO13}) to compute 
$d \Delta x/dn$ and using the 
assumed uncertainties in $n$ and $\Delta\tilde\nu$, we obtain 
\begin{equation}
\Delta R = 0.04\,\mu{\rm m}. 
\label{SRIDX6}
\end{equation}
This is consistent with the three values obtained by our 
three methods above. According to (\ref{SRIDX6}), 
the relative error in $R$ amounts to 
$\Delta R/R=0.013$, which is less than 2\%. 
Thus, while not as precise as optical methods 
for the determination of sphere radii \cite{Ashkin,Chylek3}, 
our method using FTIR in the infrared domain is quite 
adequate for biophysical applications and may be 
improved by pushing synchrotron FTIR to its current 
practical limit of $\Delta\tilde\nu\approx 1\,$cm$^{-1}$. 
 

\section{PMMA Dispersion in the Near-Infrared} 
\label{PMMAD}
In this section we determine the PMMA index of refraction 
as a function of wavenumber for our 
measured PMMA sphere. 
While we expect that this 
data is representative of PMMA,  
we should not expect that the 
fine details of this data are reproducible as a characteristic of PMMA itself. 
In fact, the precise index of refraction $n$ of PMMA depends on 
the specific batch of PMMA under investigation since there 
are natural variations depending on the production 
process. Therefore, we should not expect reproducibility between 
different batches of PMMA, but only for optical components 
produced from the same batch of PMMA. 
In \cite{Lang}, e.g., PMMA provided 
by four different suppliers showed variations 
in $n$ of about $\pm 0.006$. 
As we will see below, the refractive index extracted from 
our synchrotron data is sensitive to these variations.  
 
We derive the dispersion 
$n(\tilde\nu)$ of our measured PMMA sphere 
assuming that its radius, determined according 
to our direct method [see (\ref{SRIDX4a})], 
is $R=3.14\,\mu$m. In this 
case we may start from (\ref{THEO5}) and 
write 
\begin{equation}
x_J(n_J) = 2\pi R \tilde\nu_J , 
\label{RKNOWN1}
\end{equation}
where $x_J(n)$ is the position of the 
ripple corresponding to the magnetic mode 
with angular momentum $J$ and index of refraction $n$. 
For given $J$, 
the function $x_J(n)$ is known. It may, e.g.,  
be computed by determining the position of 
the $p=1$ pole  
of $S_m^J(x;n)$ in (\ref{THEO8}), 
or it may be determined graphically from 
wavenumber sweeps of $Q_{\rm ext}(\tilde\nu)$, 
since, although all partial waves contribute to 
$Q_{\rm ext}(\tilde\nu)$, at the position of 
a ripple with angular momentum $J$, the 
partial wave with mode number $J$ will dominate. 
In our determination of $x_J(n)$ we used the 
$Q_{\rm ext}(\tilde\nu)$-sweep method. 
 
Since we assume that $R$, $x_J(n)$, and 
$\tilde\nu_J$ 
are known, 
we may now determine the index of refraction 
of our PMMA sphere at 
$\tilde\nu_J$ by inverting (numerically), 
equation (\ref{RKNOWN1}) according to 
\begin{equation}
n_J = n(\tilde\nu_j) = x_J^{-1}(2\pi R \tilde \nu_J),
\label{RKNOWN2}
\end{equation}
where $x_J^{-1}$ is the inverse function 
of $x_J$. 
The result of (\ref{RKNOWN2}) for ripples with 
$J=7,\ldots,19$, 
listed in Table~\ref{TAB1}, is shown 
as the red filled squares in Fig.~\ref{Fig5}. 
Also shown is the result for the ripple with $J=6$ 
(magenta triangle in Fig.~\ref{Fig5}). 
This data point is shown as a different plot symbol, 
since, according to Fig.~\ref{Fig4}, it sits on a wing 
and, in addition, represents the merger of a magnetic and 
an electric ripple. This may give rise to large shifts in 
wavenumber, which make this point unreliable. 
 
\section{Discussion} 
\label{DISC}
We are not the first to use synchrotron infrared radiation 
for extinction measurements on PMMA spheres. Several 
previous measurements are 
reported in the literature 
\cite{Bassan,BPhD,Dijk}.  
However, all of these previous measurements are 
preformed in the infrared regime with wavenumbers 
$\lesssim 4000\,$cm$^{-1}$ where strong absorption 
due to chemical absorption bands occurs, which 
obscures the ripples in the absorption spectrum. 
In our measurement, in 
a deliberate attempt to access the regime of ripples, 
we extended the wavenumber range to $8000\,$cm$^{-1}$, 
above the chemical absorption regime, 
in which, for spheres with a radius of $R\lesssim 10\,\mu$m, 
the ripples are fully formed. 
We conducted FTIR spectroscopy on PMMA spheres as 
a pilot project for our ultimate goal, FTIR 
spectroscopy on cells and other microscopic, 
quasi-spherical biological 
structures, such as plant pollen. From our experiments 
with PMMA spheres we learn that FTIR spectroscopy is a powerful 
tool that, under ideal conditions, such as presented 
by PMMA spheres, is capable to determine 
radii and the dispersion of the index of refraction. 
In order to accomplish this, however, the conventional FTIR 
range, which typically ranges up to $4000\,$cm$^{-1}$, 
needs to be extended 
to at least $8000\,$cm$^{-1}$, as done in our experiments, 
in order to access 
the wavenumber regime above $4000\,$cm$^{-1}$ in which 
most organic materials, including PMMA, do not show 
chemical absorption bands. Including this wavenumber 
regime in spectroscopic sweeps, as shown 
here for PMMA, provides valuable physical information 
on the specimen under investigation. 
 
If we would like to extract information from the 
ripple structure of a scatterer, 
it is important to work with {\it isolated} scatterers. 
For instance, in 
his Ph.D. thesis \cite{BPhD}, Bassan shows 
the FTIR spectrum of a PMMA sphere that is in contact with 
and surrounded by 
at least four neighboring PMMA spheres. Although only 
the central sphere is illuminated, 
the resulting spectrum is smooth and does not show 
any ripples. This effect is understandable, since, 
as shown in Sec.~\ref{THEO}, 
the ripples are due to whispering gallery modes, and 
touching spheres spoil the whispering gallery-mode structure, 
characteristic for single, isolated spheres. 
This is so, since touching spheres are strongly coupled 
via tunneling, which results in $Q$-spoiling, an effect 
well-known 
in the field of microdisk lasers \cite{QSMDL}. 
 
Because we are scattering from a sphere, the ripples 
(resonances) are degenerate in $M$. In this case 
it is enough to specify $(J,e)$ or $(J,m)$ to 
classify individual ripples. This is what, implicitly, 
we did in Fig.~\ref{Fig1}, where, in addition 
to classifying ripples 
according to whether they are electric or magnetic, 
we use only one additional mode number, $J$, to characterize 
each ripple. 
In case the sphere is slightly deformed, has 
a rough surface, or has a non-spherical, 
inhomogeneous index of 
refraction, the single peaks in $Q_{\rm ext}$ will 
split into multiplets. In this case a different 
classification scheme has to be constructed. 
If the deformation is too large, it is well known 
from the field of quantum chaos \cite{Stoeckmann} 
that the resulting resonance structure is chaotic, 
and a straightforward classification scheme may be 
impossible in principle \cite{Gutz}. 
 
When irradiated with synchrotron light, the sphere 
itself turns into a radiation source and emits 
electromagnetic radiation. 
For wave numbers corresponding to ripples, either 
the electric or magnetic $2^J$-pole radiation is 
especially strong and causes a peak in $Q_{\rm ext}$. 
This is so, since on resonance a larger amount of light than 
in the off-resonance case is removed from the incident 
beam and re-radiated, essentially isotropically 
(especially for large $J$), in all directions. 
In case the wavelength is small and therefore the 
wave number is large, modes with relatively 
large $J$ are excited as demonstrated in Fig.~\ref{Fig3},  
where the electric field distribution of the magnetic 
$J=13$ mode is shown. $J=13$ 
corresponds to $2^{13}=8192$-pole radiation. 
It is surprising that such high-order multipole radiation 
can be resolved in our FTIR synchrotron experiments. 
 
The electromagnetic field in Fig.~\ref{Fig3} is 
computed and plotted using 
the formulas for the 
electric and magnetic fields in \cite{Newton}. 
However, we noticed a problem:  
The factors 
$\kappa$ and $k$ in the denominators of equations 
(2.122), (2.123) and (2.124), (2.125) in \cite{Newton}
are inconsistent and lead to electric and magnetic 
field modes that do not satisfy the boundary conditions. 
Deleting these factors of $\kappa$ and $k$ yields 
consistent expressions for the electric and magnetic fields, 
which we then used to plot Fig.~\ref{Fig3}. 
 
A synchrotron beam is very intense and the question of 
heating the sample under investigation may arise. 
Although a synchrotron beam is more 
intense than a beam generated by, for instance, a {\it Globar} 
infrared source \cite{Bonner}, 
the intensity is still low enough that heating of the 
sample can be completely neglected. 
This is corroborated by the photon flux per unit 
wavenumber, which is very small. 
 
Small spheres, such as the sphere measured in our 
FTIR synchrotron experiments, 
are useful for absolute calibrations. This is so, since 
for small spheres 
wiggles and ripples have a one-to-one correspondence with 
theoretical simulations of $Q_{\rm ext}$.  
Since for small spheres even the large-wavelength part of 
the spectrum is essentially undistorted by chemical absorption, 
the $J$ classification of magnetic and electric modes can 
be accomplished unambiguously. This helps greatly in determining 
both the radius of a sphere and the dispersion of the 
index of refraction. Of course, a 5-micron sphere, e.g., covers
only half the $x$ range of a 10-micron sphere. However, 
there is an upside: Precisely because only half 
the $x$ range is covered for the same 
wavenumber range, the extinction spectrum of a 
5-micron sphere has effectively twice the 
resolution than the extinction spectrum of 
a 10-micron sphere. This fact considerably facilitates 
the mode-number assignment of the spectra of 
small spheres. On the downside, the spectra of small 
spheres do not quite reach the high-$x$ regions where 
the ripples become sharp.
 
\section{Summary and Conclusions} 
\label{SUMCO}
In this paper we showed that the ripple structure in 
synchrotron FTIR extinction spectra 
is capable of determining the radius of 
a dielectric sphere with great accuracy. In addition, we showed 
that the ripple structure allows us to 
extract the frequency-dependent index of refraction
of PMMA in a wavenumber regime where it has 
not previously been measured. 
Both the determination 
of an unknown sphere radius and the determination 
of the dispersion of the index of refraction are 
applicable to any transparent material that can be 
shaped into spheres. Thus, FTIR is shown to be 
a capable method for determining the index of refraction 
in the infrared regime, needed in biophysical FTIR 
spectroscopy, but not generally available by other means. 
 
\section{Acknowledgements} 
\label{ACK}
We would like to thank the staff of the MAX III 
synchrotron in Lund, Sweden, in particular 
Anders Engdahl, 
for valuable assistance. 
Financial support by the Norwegian Science Council 
under grant number 216687, 
``Hyperspectral imaging in biophysics and energy physics'', 
is gratefully acknowledged. 
 

\end{document}